# Non-constant crack tip opening angle and negligible crack tunneling of brittle fracture in Al: A first-principles prediction

W. T. Geng[a]

*School of Materials Science & Engineering, University of Science and Technology Beijing,*

*Beijing 100083, China*

December 22, 2009

Numerous measurements showed that the crack tip opening angle (CTOA) is nearly constant upon stable ductile fracture in Al alloys which widely used in modern transportation industry. The atomic structure of the very tip of a crack front has remained unknown, however. We have carried out a first-principles density functional theory study to reveal the precise alignment of atoms near the crack tip in single-crystalline Al. The calculations demonstrate that the CTOA increases with the opening displacement, thus the observed constant CTOA in millimeter scale is an entirely plastic effect during ductile crack. Besides, we find no significant crack tunneling (crack-front blunting), which can be accounted for from the very small relaxation of the Al free surface. The atomic structure thus obtained provides a solid basis for larger scale simulations using for example finite element method.

---

[a] E-mail: geng@ustb.edu.cn

As we are now under increasingly serious threat of climate change and global warming, the demands for saving energy and reducing carbon emission are more urgent than ever. Therefore, more and more attention is being paid on the exploitation of lightweight structures such as thin-sheet Al alloys applied in aerospace and pipeline industries. To predict the failure of cracked metallic materials with the elastic-plastic finite element analyses, one of the widely used fracture criteria is the crack-tip-opening-angle (CTOA).[1,2,3] There are many measurements and numerical simulations on various alloys under low constraint suggesting that the CTOA is high in the initial stage of the crack growth and decreases progressively to a nearly constant value after several millimeters. However, a recent work combining both measurements and finite-element analyses on the crack growth in carbon steel A285 done by Lam et al. showed that under high constraint, the initial value of CTOA exerts long-range influence on the crack and can no longer be taken as constant.[4] Different from the case of brittle fracture in covalent-bonded crystals Si and SiC, where the atomic structure of the crack tip were obtained from either tight-binding or ab initio calculations,[5,6,7] the permanent plastic deformation associated with ductile cracking in metals makes the calculation of the CTOA, which is defined and measured in industrial practice in macro-scale, a formidable task from first-principles even with the most powerful supercomputer resources at the present time.

To describe ductile fracture process, Lloyd introduced the concept of a crack-tip-process-zone,[8] within which chemical bonds break to create new surfaces and dislocations emanate, and out of which elasticity dominates the deformation. Obviously, this crack-tip-process-zone is hard to precisely define, but it should typically span multi grains along the crack path. To understand the structural evolution during the crack growth in micro-scale, we need to figure out first the atomic structure of the crack tip. In polycrystalline materials, the tip of a crack is either inside a grain or on the grain boundary. When the grain size is very small, dislocation emission and movement can hardly occur. Atomistic simulations on nano-crystalline Ni with mean grain size ranging from 5 to 12 nanometers suggest that intergranular fracture is dominant.[9] But here for Al, we take the other cases for consideration: the grain sizes are well above nanoscale and the crack tip has a good chance to be inside a grain. If we take the average dislocation density of Al as $10^8$-$10^9$/$m^2$ [Ref. [10]], it is reasonable to assume that on average the dislocations are over dozens of nanometers away from the crack tip. Due to the nearsightedness of electrons in condensed matter,[11] such a distance is large enough to minimize the impact of dislocations on the atomic arrangements at the crack tip. It follows that the brittle fracture of a perfect crystalline Al, which happens at the very front of the tip, is indeed relevant to the whole ductile fracture process in a polycrystalline Al. Consequently, a scrutiny of the brittle fracture of a perfect crystalline Al from quantum mechanical view could serve as a good starting point for attacking the complicated problem, a solid basis for some inputs employed in finite-element simulations.[12] With this aim in mind, we set out to calculate the structural and electronic properties of a crack tip in a stable brittle tearing process in single crystalline Al.

As we did in the study of grain boundary cohesion in transition metals,[13,14] we have used a slab to model a stable crack in Al (Fig. 1). We assume the tearing is along the (100) plane, [110] direction. There are 17 layers in the [001] direction, and 16 or 17 atoms in each layer. The calculated lattice constant for Al is 4.04 Å, in excellent agreement with experimental value.[15] Thus, the dimensions of the crack-free slab are about 2.86 Å ×32 Å ×40 Å. With a periodic boundary condition, we have separated the neighboring slabs in both [001] and [110] directions by a vacuum region of at least 10 Å to minimize the interaction between neighboring slabs. Atoms on the left side of the slab were fixed to the corresponding positions in a free (001) surface. The right-most atoms on both top and bottom surfaces were fixed in [001] direction in order to keep the designed crack opening displacement (height of the crack at the right end of the slab); whereas their freedom in the [110] direction was optimized. All the other atoms were fully relaxed to reach a stable tearing state. We note that with such an initial setup, we were in fact dealing with a healing crack. The critical load thus determined is indeed the lower limit if there is crack tunneling (i.e., tip blunting, also termed as *lattice trapping*) at the tip. To ensure a realistic description of the very tip of a crack, we have also tested the convergence of CTOA on the length in *a* direction by removing two surface atomic layers on the left end of the slab. For a real specimen, the three-dimensional crack resistance is dependent on all the dimensions *a*, *b*, and *c*. Here we are dealing with planar stresses, thus dimension *b* is not a concern.

**Fig. 1. (Color online) Slab model used to simulate a healing crack in face-centered-cubic Al. Large and small circles represent Al atoms in different (110) plane. Atoms on the left side of the slabs were fixed to the corresponding positions in a free Al (001) surface. The right-most atoms on both top and bottom surfaces were fixed in [001] direction in order to keep the designed crack opening displacement, which is 15Å in (a) and 10Å in (b).**

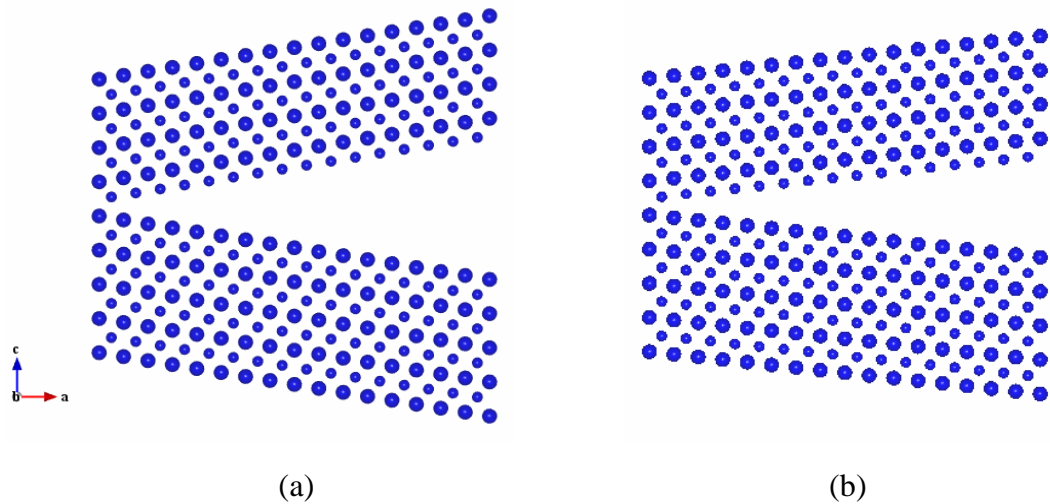

(a) (b)

The first-principles density functional theory calculations were conducted using Vienna *Ab initio* Simulation Package (VASP).[16] The electron-ion interaction was described using projector augmented wave (PAW) method.[17] The exchange correlation between electrons was treated with generalized gradient approximation (GGA) in the Perdew-Burke-Ernzerhof (PBE) form.[18] The energy cutoff for the plane wave basis set was 240 eV for all calculated systems. The Brillouin-zone integration was performed within Monkhorst-Pack scheme using four $k$ points and the Methfessel-Paxton smearing with a width of 0.02 eV. The energy relaxation for each strain step is continued until the forces on all the atoms are converged to less than $3\times10^{-3}$ eV Å$^{-1}$.

Figure 2 displays the optimized geometry of the slab shown in Fig. 1. The most salient feature of the crack tip in single-crystalline Al during stable healing is that there is no significant crack tunneling, or lattice trapping. In fact, the CTOA is very well defined for both crack openings. This is in sharp contrast to the case of Si (Ref. 6 and 7) where crack tunneling is significant due to directionality of the covalent chemical bonds. Moreover, we find that the CTOA varies with the crack opening. It is about 29° for a crack opening of 15 Å (left) and 23° for 10 Å (right). Starting from the optimized configuration in panel b, we removed the left-most two atomic layers and then fix the third layer before we re-optimize the geometry. It turns out that although some atoms around top-left and down-left corners experienced some relaxation, the CTOA is hardly changed, indicating the left end of the specimen is already far enough away from the crack tip. On the other hand, removal of surface layers in c direction does enlarge the CTOA due to reduction in bending (flexural) resistance; whereas the crack tunneling remains negligible.

**Fig. 2. (Color online) The optimized geometry of the slabs depicted in Figure 1.**

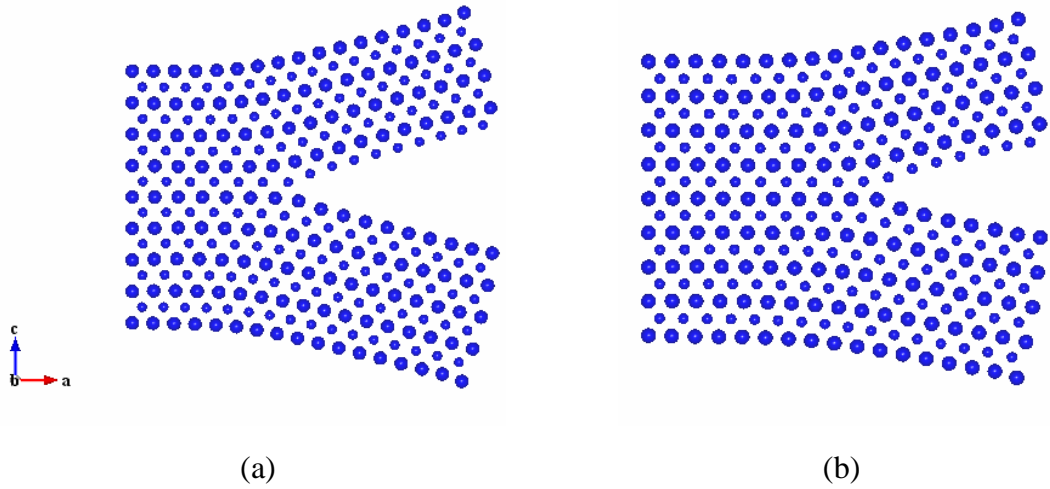

(a)                  (b)

To make the lattice distortion near the crack tip more visible, we draw in Fig. 3 the calculated valence charge density of the two configurations shown in Fig. 2. Bending of these two brunches can be easily seen from the electron accumulation in the center part of the top and down surfaces and electron depletion around the crack tip. Away from the surface and crack, the charge density recovers to the bulk value.

**Fig. 3. (Color online) The calculated valence charge density for the systems displayed in Fig. 2.**

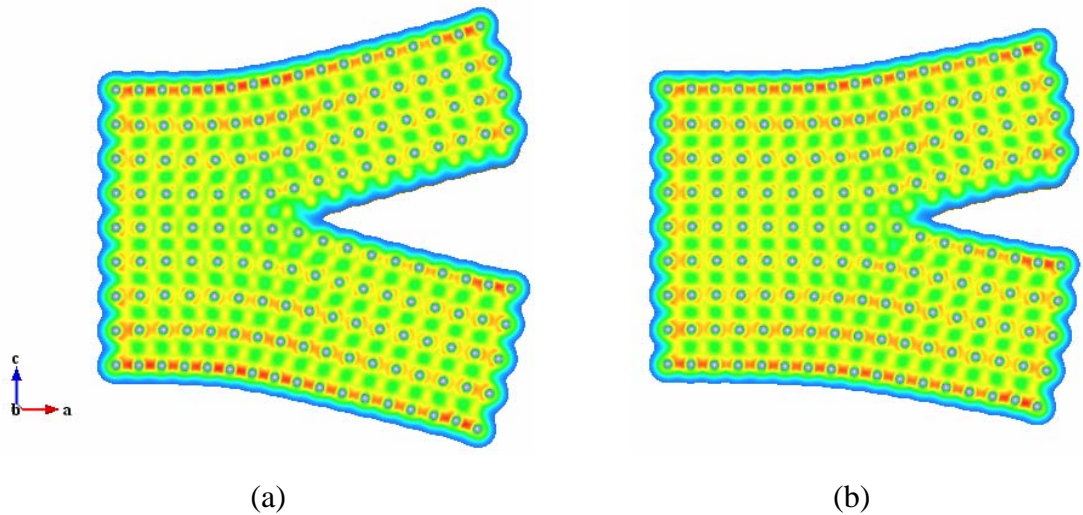

(a)                  (b)

According to the Griffith criterion,[19] for perfect brittle fracture, the gain in elastic energy associated with the crack advance, $E_e(\theta)$, is twice the decrease in surface energy, $E_s(\theta)$, which are all functions of the CTOA, $\theta$. In equilibrium, the driving forces for tearing and healing of the crack are canceled exactly to each other. That is,

$$\frac{\partial E_e(\theta)}{\partial \theta} = \frac{\partial E_s(\theta)}{\partial \theta}. \qquad (1)$$

It follows immediately that if the cleavage energy is reduced, so will be the CTOA. Since to the best of our knowledge there have been no report on first-principles quantum mechanical study of the CTOA in metals, we have conducted full calculations to examine the effect of variation of cleavage energy on the CTOA, in order to gain confidence in our numerical results. We removed the downside interfacial atomic layer of the specimen showed in Fig. 2a to form a dislocation loop. Apparently, cleavage becomes easier as the chemical bonding across the loop is weakened due to loss of some nearest neighbors. Our calculations confirm this expectation (Fig. 4a). The CTOA reduced from 29° to 24°. Moreover, we also examined the change in CTOA when two atomic layers are removed from both the top and bottom surfaces. As the specimen gets thinner, bending is easier. That is, $\frac{\partial E_e(\theta)}{\partial \theta}$ becomes smaller. Our calculation shows the CTOA increases from 29° to 33°.

**Fig. 4. (Color online) (a) The optimized atomic configurations when one interfacial atomic layer is removed from the slab depicted in Fig. 2a, (b) or when two atomic layers are removed from both the top and bottom surfaces.**

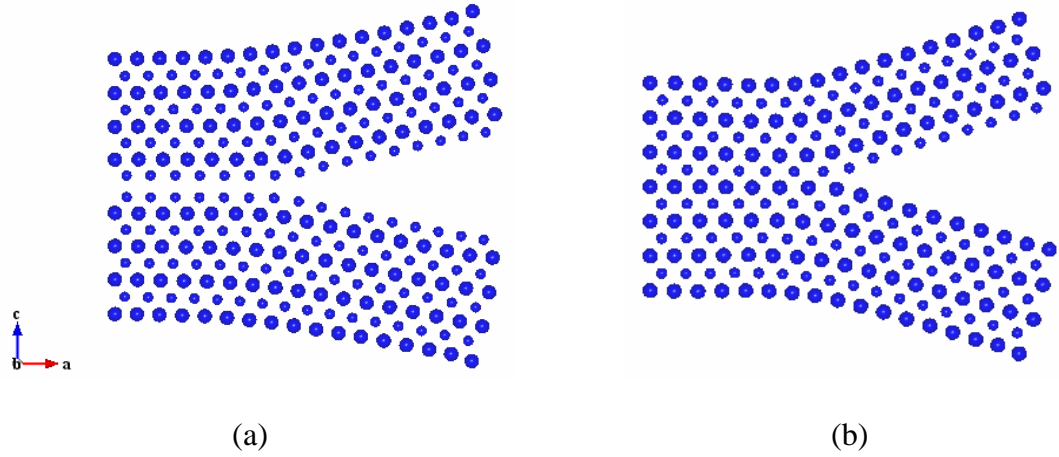

(a)            (b)

With the confidence of our computation model, it appears safe to draw the conclusion that the CTOA is not constant during stable tearing of single-crystalline Al. As a consequence, the observed constant CTOA in industrial thin-sheet Al alloys is completely a plastic effect. It is certainly desirable to develop a quantitative model to predict the value of CTOA for a material. Such an attempt is indeed a challenging task, because the analytical form for the elastic energy, which is related directly to the bending modulus, is difficult to reach. Bending modulus is defined as

$$\varepsilon_b = \frac{Fl^3}{4wdt^3}, \qquad (2)$$

where $F$ is the normal force acted on the beam, $l$, $w$, and $t$ are the length, width, thickness of the beam, and $d$ is deflection at the load point. Only for small deflections can it be simply related to the Young's modulus. On the other hand, if there is crack tunneling (or, if the crack tip is not atomically sharp), the area of surface formed following the crack will not be well defined in the atomic scale. Equation (1) will not hold for a state in equilibrium.

The reason that we find no noticeable crack tunneling for Al is that the multilayer relaxation in Al (001) is very small, unlike the case of high index Al surface.[20] According to our calculation, the contraction of the outermost interlayer distance in Al(001) is only about 0.01Å. For other transitional metals such as Fe,[21] Ni,[22] and Cu,[23] both experimental and theoretical works reveal remarkable (> 0.1 Å) contraction in the surface interlayer distance. As a result, the crack tip in those metals will be blunted in atomic scale, resulting in a reduction of the fracture resistance.

To summarize, we have carried out a density functional theory study on the precise alignment of atoms near the crack tip in single-crystalline Al. We find that the crack tip opening angle (CTOA) increases with the opening displacement. Therefore, the observed constant CTOA in industrial thin-sheet of Al in millimeter scale is an entirely plastic effect during ductile crack. Furthermore, we find no noticeable crack tunneling. These numerical predictions, if confirmed by microscopic measurements, could be useful parameters for larger scale simulations on ductile crack.

I am grateful to the support of NCET (Grant No. 06-0080). The calculations were carried out on the Quantum Materials Simulator of USTB.